\documentclass{ws-ijmpb-rev}

\begin{document}

\markboth{N.-C. Yeh and C.-T. Chen}
{Non-Universal Pairing Symmetry and Pseudogap Phenomena ...}

%
%

\title{NON-UNIVERSAL PAIRING SYMMETRY AND PSEUDOGAP\\
PHENOMENA IN HOLE- AND ELECTRON-DOPED\\
CUPRATE SUPERCONDUCTORS}

\author{N.-C. YEH\footnote{Corresponding author. E-mail: ncyeh@caltech.edu} \ and C.-T. CHEN}

\address{Department of Physics, California Institute of Technology\\
Pasadena, CA 91125, USA}

\maketitle


\begin{abstract}
Experimental studies of the pairing state of cuprate
superconductors reveal asymmetric behaviors of the hole-doped
(p-type) and electron-doped (n-type) cuprates. The pairing
symmetry, pseudogap phenomenon, low-energy spin excitations and
the spatial homogeneity of the superconducting order parameter
appear to be non-universal among the cuprates, which may be
attributed to competing orders. We propose that the non-universal
pseudogap and nano-scale variations in the quasiparticle spectra
may be the result of a charge nematic (CN) phase stabilized by
disorder in highly two-dimensional (2D) p-type cuprates. The CN
phase is accompanied by \textsl{gapped} spin excitations and
competes with superconductivity (SC). In contrast,
\textsl{gapless} spin excitations may be responsible for the
absence of pseudogap and the presence of excess sub-gap spectral
weight in the momentum-independent quasiparticle spectra of n-type
cuprates. The physical implications and further verifications for
these conjectures are discussed.
\end{abstract}


\section{Introduction}

Recent spectroscopic studies of various hole-doped (p-type) and
electron-doped (n-type) cuprate superconductors have revealed that
some phenomena widely perceived as essential for the occurrence of
cuprate superconductivity are in fact not universal, and that
substantial asymmetries exist between p-type and n-type
cuprates.\cite{1,2,3,4} More specifically, the pairing symmetry in
p-type cuprates appears to be predominantly $d_{x^2-y^2}$ or
$(d_{x^2-y^2}+s)$ for all doping levels, both involving four nodes
in the momentum ($k$) space and $k$-dependent superconducting
energy gap ($\Delta _k$).\cite{1,2,3} In contrast, quasiparticle
tunneling spectra of infinite-layer and one-layer n-type cuprates
studied to date are all $k$-independent and are therefore
consistent with $s$-wave pairing,\cite{3,4,5} although experiments
other than tunneling spectra suggest doping dependent pairing
symmetry in the one-layer n-type cuprates.\cite{6,7,8} These
different pairing symmetries have been attributed to competing
energy scales in the cuprates.\cite{1} The response in the pairing
state to quantum impurities that substitute Cu in the CuO$_2$
plane is also asymmetric between p-type and n-type cuprates, the
former being sensitive to both magnetic and non-magnetic
impurities as expected for $d$-wave
superconductors,\cite{2,3,9,10} while the latter being only
sensitive to magnetic impurities as expected for typical $s$-wave
superconductors.\cite{4} Moreover, scanning tunneling and angular
resolved photoemission spectroscopic (STS and ARPES)
studies\cite{11,12} reveal pseudogap features well above the
superconducting transition T$_c$ in optimal and underdoped $\rm
Bi_2Sr_2CaCu_2O_{8+x}$ (Bi-2212) single crystals, and these
pseudogap features evolve smoothly  below $T_c$into the so-called
``dip-hump'' satellite features at energies above the
superconducting gap. In contrast, no quasiparticle pseudogap has
been identified in $\rm YBa_2Cu_3O_{7-\delta}$ (YBCO) above
$T_c$,\cite{13} although satellite features similar to those of
Bi-2212 also exist at $T < T_c$.~\cite{2,3,13} On the other hand,
neither spectral satellite features below $T_c$ nor quasiparticle
pseudogap above $T_c$ have ever been reported for the n-type
cuprates.\cite{4,15,16} The low-energy spin excitations inferred
from neutron scattering studies also reveal substantial
differences between the p-type and n-type cuprates, the former
being associated with gapped spin excitations\cite{17} while the
latter being gapless.\cite{18,19} Finally, nano-scale spatial
variations in the tunneling gap of under- and optimally doped
Bi-2212\cite{20} differ from the spatially homogeneous pairing
potential in YBCO and n-type cuprates according to STS\cite{2,3,4}
and nuclear magnetic resonance (NMR) studies.\cite{21}

The aforementioned non-universal phenomena strongly suggest that
the asymmetric behaviors between n-type and p-type cuprates cannot
be easily reconciled with a one-band Hubbard model that asserts
particle-hole symmetry,\cite{1} and that better understanding of
the pairing mechanism relies on determining the physical origin of
these non-universal phenomena and identifying the ubiquitous
features among all cuprates. In the following, we propose a
scenario that provides feasible account for the non-universal
pseudogap phenomenon, satellite features and spatial homogeneity
of the pairing potential. The physical implications of and more
stringent tests for this scenario are discussed.

\section{Pseudogap Phenomena, Spin Excitations and Spatial Homogeneity of the Order Parameter}

The absence of pseudogap in n-type cuprates\cite{4,15,16} implies
that the pseudogap phenomenon cannot be a prerequisite for cuprate
superconductivity. To better understand the physical origin of the
pseudogap, we compare the quasiparticle tunneling spectra in
under- and optimally doped Bi-2212 with those of YBCO and n-type
cuprates, as exemplified in Figs. 1(a)-(b). In the case of
Bi-2212, a typical c-axis tunneling spectrum with sharp
superconducting coherence peaks at $T < T_c$ is often accompanied
by the dip-hump satellite features at higher energies,\cite{11,20}
and the superconducting coherence peaks diminish with increasing
temperature.\cite{11} The other types of spectra in Bi-2212, which
typically occur at locations only a few nanometers away from
regions showing the first type of spectra,\cite{20} reveal rounded
peaks and non-vanishing density of states at zero bias, as
exemplified in the inset of Fig. 1(a), and these rounded features
generally persist well above $T_c$.\cite{11} In contrast, only one
type of c-axis tunneling spectra with both sharp coherence peaks
and satellite features can be observed in YBCO,\cite{1,2,3,13,14}
as shown in the main panel of Fig. 1(a), and all interesting
features completely vanish above $T_c$.\cite{13}

\begin{figure}[th]
\centerline{\psfig{file=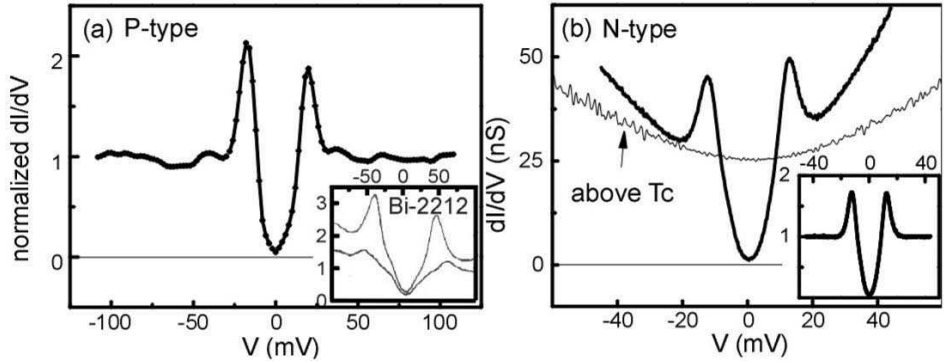,width=12.5cm}} \vspace*{8pt}
\caption{Comparison of the quasiparticle tunneling spectra between
p-type and n-type cuprates. (a) c-axis tunneling spectrum of an
optimally doped YBCO at 4.2 K.$^{2,3}$ The inset shows sketches of
two typical types of c-axis tunneling spectra of a slightly
underdoped Bi-2212.$^{11,20}$ (b) Momentum-independent tunneling
spectra of an infinite-layer n-type cuprate $\rm
Sr_{0.9}La_{0.1}CuO_2$ at 4.2 K $\ll T_c = 43$ K and above
$T_c$.$^{3,4}$ The inset shows the normalized spectrum of the main
panel.}
\end{figure}

The phenomena described above may be understood in terms of a
``pseudogap phase'' competing with superconductivity as follows.
Theoretical investigations\cite{22,23} have shown that two
competing orders A and B of a system generally result in three
types of phase diagrams as a function of the chemical potential
$\mu$, as illustrated in Figs. 2(a)-(c). For the phase diagram
depicted in Fig. 2(a), nano-scale phase separations of A and B are
expected if the chemical potential coincides with the critical
value $\mu _c$. Noting that the same chemical potential can be
associated with different doping levels, we suggest that the
under- and optimally doped Bi-2212 may belong to this category. On
the other hand, long-range spatial homogeneity of the order
parameter is expected for the coexisting phase A/B depicted in
Fig. 2(b). This scenario may correspond to the observed spatial
homogeneity of the quasiparticle spectra in the pairing state of
YBCO, with coexisting superconducting coherence peaks and
satellite features. Moreover, it could account for the absence of
quasiparticle pseudogap above $T_c$\cite{13} if the condition $\mu
_x \le \mu \le \mu _b$ is satisfied.

\begin{figure}[th]
\centerline{\psfig{file=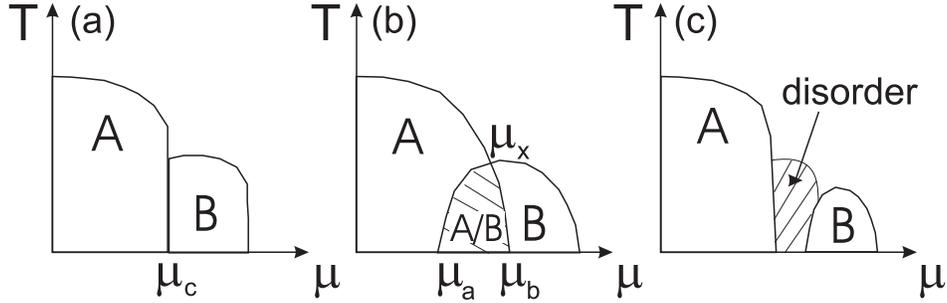,width=12.5cm}} \vspace*{8pt}
\caption{A schematic illustration of three possible temperature
($T$) vs. chemical potential ($\mu$) phase diagrams associated
with two competing orders A and B, depending on details of the
Hamiltonian.$^{22,23}$ (a) A and B are separated by a first-order
critical line or critical point. (b) A and B coexist over a finite
range of chemical potential, and the coexisting phase A/B is
separated from A and B with second-order critical lines. (c) A and
B are intervened by disorder. In p-type cuprates, A may be a
charge nematic phase associated with gapped spin excitations, and
B may correspond to the superconducting phase.}
\end{figure}

Concerning the microscopic makeup of the pseudogap phase, we note
that according to the Mermin-Wagner theorem,\cite{24} no
long-range order with spontaneous breaking of continuous symmetry
can exist in a strictly 2D system. Given that superconductivity
involves spontaneous U(1) symmetry-breaking below $T_c$, one would
expect the absence of long-range superconducting order in highly
2D systems such as Bi-2212. Meanwhile, if a second phase
relatively close in energy to superconductivity (SC) exists and if
it can be stabilized by disorder, phase separations at between the
second phase and SC at relatively small spatial scales would
become highly probable. We suggest that the pseudogap phase may be
a ``charge nematic'' (CN) phase which involves localized
stripe-like charge modulations. The CN phase is energetically more
favorable than charge stripes because of its reduced Coulomb
repulsion relative to the latter. Moreover, CN phase could fulfill
the role of accommodating incommensurate hole doping via local
spin fluctuations, which is consistent with neutron scattering
experiments that attribute the observation of gapped spin
excitations to charge modulations in p-type cuprates.\cite{17}
Under this premise, the phase boundary between CN and SC would
involve defected spin configurations because SC is effectively a
spin liquid, and the defected spin configurations in 2D could be
either positive or negative in sense, similar to the disclinations
or vortex/anti-vortex pairs in 2D. Moreover, these locally
defected configurations could be statically pinned by disorder,
analogous to the observed quantum Hall nematic (QHN) phase in 2D
electron gas (2DEG).\cite{25} On the other hand, introduction of
3D coupling not only stabilizes long-range SC order but also
permits smooth relaxation of defected spin configurations, thereby
eliminating the need for phase separation. Thus, our conjecture of
CN being the pseudogap phase competing with SC can account for the
long-range spatial homogeneity of the pairing potential in 3D YBCO
and the nano-scale phase separations in 2D Bi-2212 for $T < T_c$.
In contrast, gapless spin excitations in n-type
cuprates\cite{18,19} are unlikely related to local charge
modulations. Therefore there is no obvious competing order with SC
near the optimal doping of n-type cuprates, which explains the
long-range homogeneity of the pairing potential and the absence of
pseudogap in zero field.\cite{1,4,15,16}

\section{Spectroscopic Evidence for Gapless Spin Excitations in N-Type Cuprates}

In the case of n-type cuprates, while the momentum-independent
quasiparticle tunneling spectra and the insensitivity of
superconductivity to non-magnetic impurities are strongly
supportive of s-wave pairing symmetry,\cite{4,26} the
quasiparticle spectra differ substantially from the BCS prediction
for conventional s-wave superconductors.\cite{4} As illustrated in
Fig. 3, we note that there are excess sub-gap quasiparticle
density of states (DOS), which cannot be accounted for by either
disorder effects\cite{4} or anisotropy in the fully gapped Fermi
surface.\cite{26} Recall that neutron scattering experiments
reveal gapless spin excitations in n-type cuprates,\cite{18,19} we
conjecture that the excess sub-gap quasiparticle DOS is the result
of quasiparticle coupling to the background gapless spin
excitations. However, theoretically it is not possible to have
gapless spin excitations above one dimension without the
spontaneous breaking of a continuous symmetry.\cite{24} In the
absence of spontaneous SU(2) symmetry-breaking in the
superconducting state, the only possibility for generating gapless
spin excitations in n-type cuprates is to invoke non-trivial
coupling of spin fluctuations to the phase of the superconducting
order parameter. The resulting phase fluctuations have been shown
to yield broadening of the quasiparticle spectra,\cite{27} which
is consistent with our experimental finding.\cite{4} Such coupling
can also justify the existence of gapless spin excitations through
spontaneous breaking of the U(1) symmetry-breaking in the
superconducting state. However, a rigorous proof from first
principle will require deriving an effective perturbation
Hamiltonian ${\cal H} _1$ from the coupling of the gapless spin
excitations to the phase of the superconducting order parameter.
Once ${\cal H} _1$ is known, one can obtain the Green's function
${\cal G}$ of the perturbed system, and the quasiparticle DOS
${\cal N} (E)$ can be derived through the imaginary part of ${\cal
G}$ via the relation ${\cal N} (E) = - {\rm Im} [{\cal G}] / \pi$.
This theoretical issue awaits further investigation.

\begin{figure}[th]
\centerline{\psfig{file=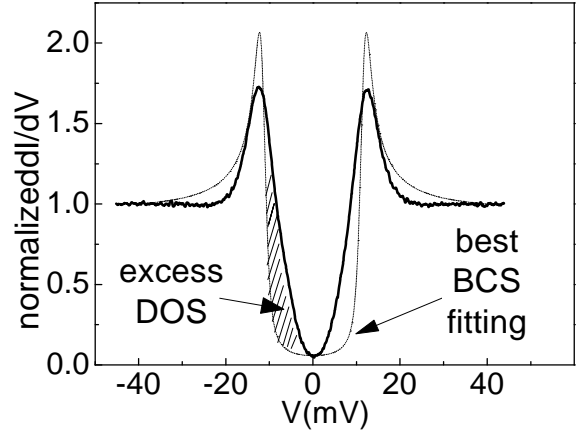,width=7.5cm}}
\vspace*{8pt}\caption{Comparison of the normalize quasiparticle
tunneling spectrum of the infinite-layer n-type cuprate
Sr$_{0.9}$La$_{0.1}$CuO$_2$ with the best BCS fitting at 4.2 K,
showing apparent disagreement and excess sub-gap quasiparticle DOS
relative to conventional BCS theory.$^{4}$ We attribute the excess
sub-gap DOS to quasiparticles damping by the gapless spin
excitations in the system.}
\end{figure}

\section{Discussion}

We have proposed in the previous sections that different
low-energy spin excitations in p-type and n-type cuprate
superconductors are responsible for various non-universal
characteristics such as the pseudogap phenomenon and satellite
features in the quasiparticle spectra. We further suggest that a
charge nematic (CN) phase as a competing order with
superconductivity can gives rise to nano-scale spectral variations
in highly 2D p-type cuprates while yielding long-range spectral
homogeneity in 3D p-type cuprates. The conjecture of a competing
CN phase is feasible for a number of reasons. First, CN is much
closer in energy to SC than other known competing orders such as
antiferromagnetism and stripes. Moreover, the association of the
CN phase with gapped spin excitations is consistent with the
doping dependence of pseudogap in p-type cuprates, because the
spin stiffness and therefore the energy required for gapped spin
excitations decreases with increasing hole doping. In addition, CN
as a near-by competing order would be consistent with the
``checker-board'' short-range charge modulations observed within
the vortex core of Bi-2212\cite{28,29} upon the suppression of SC
inside the vortex core by magnetic fields. Finally, the absence of
gapped spin excitations in n-type cuprates implies absence of a
competing CN phase, which can be reconciled with the long-range
spectral homogeneity in n-type cuprate as well as the absence of
pseudogap above $T_c$ and missing satellite features below $T_c$.

Despite the consistencies, a number of issues still await further
verifications. First, rigorous theoretical proof for the stability
of a ground state CN phase and its dependence on the doping level
and disorder will be necessary. In addition, the evolution of the
CN phase with temperature, the spatial distribution of the CN and
SC phase boundaries, and the effect of magnetic fields are issues
to be addressed both empirically and theoretically. Finally, the
microscopic coupling mechanism of the gapless spin excitations in
n-type cuprates to phase fluctuations of the order parameter must
be resolved.

\section{Summary}

We have reviewed asymmetric behaviors among p-type and n-type
cuprates and discussed possible physical origin for various
non-universal phenomena. In the case of p-type cuprates, we
propose a charge nematic (CN) phase, which is associated with
gapped spin excitations, as a competing order to superconductivity
(SC). The competition of CN and SC can result in nano-scale phase
separation in 2D while yielding homogeneous order parameter in 3D.
On the other hand, we suggest that the gapless spin excitations in
n-type cuprates may be responsible for the excess sub-gap
quasiparticle density of states and the missing satellite features
below $T_c$, as well as the absence of pseudogap above $T_c$.
These considerations of empirical facts suggest that spin
correlation plays an essential role in cuprate superconductivity.

\section*{Acknowledgments}

This work is supported by the National Science Foundation through Grant \#DMR-0103045.

\end{document}